\documentclass[fleqn,10pt]{wlscirep}
\usepackage[utf8]{inputenc}
\usepackage[T1]{fontenc}
\usepackage{bm}
\title{The role of gain neuromodulation in layer-5 pyramidal neurons}

\author[1,*]{Alejandro Rodriguez-Garcia}
\author[2]{Christopher J. Whyte}
\author[2]{Brandon R. Munn}
\author[3,4,5]{Jie Mei}
\author[2]{James M. Shine}
\author[1,6,*]{Srikanth Ramaswamy}

\affil[1]{Neural Circuits Laboratory, Biosciences Institute, Faculty of Medical Sciences, Newcastle University, Newcastle upon Tyne, United Kingdom}
\affil[2]{Brain and Mind Center, The University of Sydney, Sydney, Australia, Center for Complex Systems, The University of Sydney, Sydney, Australia}
\affil[3]{IT:U Interdisciplinary Transformation University Austria, Linz, Austria}
\affil[4]{International Research Center for Neurointelligence, The University of Tokyo, Tokyo, Japan}
\affil[5]{Department of Anatomy, University of Quebec in Trois-Rivieres, Trois-Rivieres, QC, Canada}
\affil[6]{Theoretical Sciences Visiting Program (TSVP), Okinawa Institute of Science and Technology Graduate University, Okinawa, Japan}

\affil[*]{Correspondence: \href{mailto:a.rodriguez-garcia2@newcastle.ac.uk}{a.rodriguez-garcia2@newcastle.ac.uk}, \href{mailto:srikanth.ramaswamy@newcastle.ac.uk}{srikanth.ramaswamy@newcastle.ac.uk}}

\begin{abstract}
Biological and artificial learning systems alike confront the plasticity–stability dilemma. In the brain, neuromodulators such as acetylcholine and noradrenaline relieve this tension by tuning neuronal gain and inhibitory gating, balancing segregation and integration of circuits. Fed by dense cholinergic and noradrenergic projections from the ascending arousal system, layer-5 pyramidal neurons in the cerebral cortex offer a relevant substrate for understanding these dynamics. When distal dendritic signals coincide with back‑propagating action potentials, calcium plateaus turn a single somatic spike into a high‑gain burst, and interneuron inhibition sculpts the output. These properties make layer-5 cells gain‑tunable amplifiers that translate neuromodulatory cues into flexible cortical activity. To capture this mechanism we developed a two-compartment Izhikevich model for pyramidal neurons and single‑compartment somatostatin (SOM) and parvalbumin (PV) interneurons, linked by Gaussian connectivity and spike-timing-dependent plasticity (STDP). The soma and apical dendrite are so coupled that somatic spikes back‑propagate, while dendritic plateaus can switch the soma from regular firing to bursting by shifting reset and adaptation variables.  We show that stronger dendritic drive or tighter coupling raise gain by increasing the likelihood of calcium‑triggered somatic bursts. In contrast, dendritic-targeted inhibition suppresses gain, while somatic-targeted inhibition raises the firing threshold of neighboring neurons, thus gating neurons output. Notably, bursting accelerates STDP, supporting rapid synaptic reconfiguration and flexibility.This suggests that brief gain pulses driven by neuromodulators could serve as an adaptive two-timescale optimization mechanism, effectively modulating the synaptic weight updates.
\end{abstract}
\begin{document}

\flushbottom
\maketitle

\thispagestyle{empty}

\section{Introduction}

Layer 5 pyramidal neurons exhibit backpropagation-activated calcium (BAC) firing, where coincident distal dendritic inputs and somatic backpropagating action potentials (BAPs) trigger dendritic $Ca^{2+}$ spikes \cite{larkum_new_1999}. This shifts neuronal output from isolated spikes to bursts, encoding somatic-dendritic coincidence and modulating neuronal gain \cite{larkum_top-down_2004}. Such gain modulation is fundamental to a variety of physiologically relevant behaviors and adaptive cognitive processes \cite{salinas_correlated_2001}, including attentional shifts and perceptual switching \cite{wainstein_gain_2025, whyte_burst-dependent_2025}.\\
The ascending arousal system, particularly through noradrenergic (NA) and cholinergic (ACh) modulation, dynamically regulates neuronal activity during cognitive state transitions while preserving overall network stability \cite{munn_neuronal_2023, shine_neuromodulatory_2023}. These neuromodulators influence both the strength of apical dendritic inputs and the coupling between apical dendrites and the soma, effectively reducing the amplitude of $Ca^{2+}$ signals reaching the soma \cite{samuels_functional_2008, labarrera_adrenergic_2018, williams_dendritic_2019}. In this line, recent studies highlight how shifts in arousal states, mediated by NA and ACh, modulate the neural circuit's energy landscape by flattening or deepening it \cite{munn_neuronal_2023, wainstein_gain_2025, shine_neuromodulatory_2023}, providing a powerful framework for neuromodulatory learning independent of third-factor rules. Here, we investigate the interplay between gain modulation and spike-timing-dependent plasticity (STDP) learning within a biophysically realistic network composed of layer-5 pyramidal neurons interacting with dendritic-targeting somatostatin (SOM) interneurons and somatic-targeting parvalbumin (PV) interneurons.\\
\section{Methods}

\subsection{Network level}
The layer-5 cortical spiking neural-network model was developed to capture the neuromodulatory gain-control mechanism. We considered a recurrent network composed of 80 \% pyramidal neurons, 10 \% somatic-targeting PV interneurons, and 10 \% dendritic-targeting SOM interneurons. The network topology comprises $N^2 (N=10)$ neurons arranged on a two-dimensional $N\times N$ grid with periodic boundary conditions (a torus), so that each neuron has neighbors in all directions without edges (Figure \ref{fig:diagram}c). The synaptic weights were initialized following a Gaussian connectivity profile \cite{munn_neuronal_2023, munn_thalamocortical_2023,whyte_burst-dependent_2025} as
\begin{equation}
w_{j \to i} =
\begin{cases}
+ \; C_{E} \exp\biggl(-\dfrac{r_{ij}^{2}}{d_{E}}\biggr), & j \in \{\mathrm{Pyr}\},\\[1em]
- \; C_{I} \exp\biggl(-\dfrac{r_{ij}^{2}}{2\,d_{I}^{2}}\biggr), & j \in \{\mathrm{PV},\,\mathrm{SOM}\}.
\end{cases}
\end{equation}
where $r_{ij}$ is the Euclidean distance between neuron $i$ and $j$, $C_E$ and $C_I$ set the peak amplitude of excitation and inhibition and $d_E$ and $d_I$ their spatial extents. To respect the known compartmental targeting of cortical interneurons, we restricted SOM connections to inhibit only the apical dendrites in layer I (dendritic-targeting interneurons, DTIs), confined PV connectivity to the somatic compartment in layer V (somatic-targeting interneurons, STIs), and allowed pyramidal neurons to form recurrent excitatory synapses across their full spatial extent, reflecting their long-range projections(Figure \ref{fig:diagram}a). The network plasticity evolved according to the spike-timing-dependent plasticity (STDP) rule with eligibility traces \cite{florian_reinforcement_2007}, i.e.
\begin{align}
\frac{dP^{+}_{ij}}{dt}
&= -\frac{P^{+}_{ij}}{\tau_{+}}
   + A_{+}\sum_{k}\delta\bigl(t - t_{i}^{(k)}\bigr), \\[1ex]
\frac{dP^{-}_{ij}}{dt}
&= -\frac{P^{-}_{ij}}{\tau_{-}}
   + A_{-}\sum_{m}\delta\bigl(t - t_{j}^{(m)}\bigr), \\[1ex]
\frac{dz_{ij}}{dt}
&= -\frac{z_{ij}}{\tau_{z}}
   + \frac{1}{\tau_{z}}\Bigl[
       P^{+}_{ij}\sum_{m}\delta\bigl(t - t_{j}^{(m)}\bigr)
       + P^{-}_{ij}\sum_{k}\delta\bigl(t - t_{i}^{(k)}\bigr)
     \Bigr], \\[1ex]
\frac{dw_{ij}}{dt}
&= \eta\,z_{ij}.
\end{align}
where $t_i^{(k)}$ and $t_j^{(m)}$ denote the $k$-th presynaptic and $m$-th postsynaptic spike times, respectively; $\tau_{+}$, $\tau_{-}$ set the STDP window time decays, with $A_{+}$ and $A_{-}$ the LTP/LTD amplitudes; $\tau_{z}$ is the eligibility‐trace decay constant; and $\eta$ the learning rate. At each integration step these updates are applied and then the absolute weight is clipped to $[w_{\min},,w_{\max}]$ in order to maintain numercial stability.\\
Network activity was generated by simulating an external thalamic drive to both the apical and somatic compartments of each cell. In practice, for each neuron this drive was modelled as an Ornstein–Uhlenbeck process \cite{whyte_burst-dependent_2025} with time constant $\tau_\mathrm{OU}$ and noise amplitude $\sigma$, i.e.
\begin{equation}
dI_{e,i}(t)
=\frac{1}{\tau_{\mathrm{OU}}}\bigl(\mu_i - I_{e,i}(t)\bigr)\,dt
+\sigma\sqrt{\frac{2}{\tau_{\mathrm{OU}}}}\,dW_i(t).
\end{equation}
were $dW_i$ is a standard Wiener increment introducing Gaussian noise to capture the rapid, fluctuating nature of thalamic input currents. Specifically we applied different mean currents for each compartment and cell type: $\mu_i = \{ I_e^\mathrm{dend,Pyr}, I_e^\mathrm{soma,Pyr}, I_e^\mathrm{soma,PV}, I_e^\mathrm{soma,SOM} \}$.

\subsection{Celular level}
\begin{figure*}[ht]
    \centering
    \includegraphics[width=\textwidth]{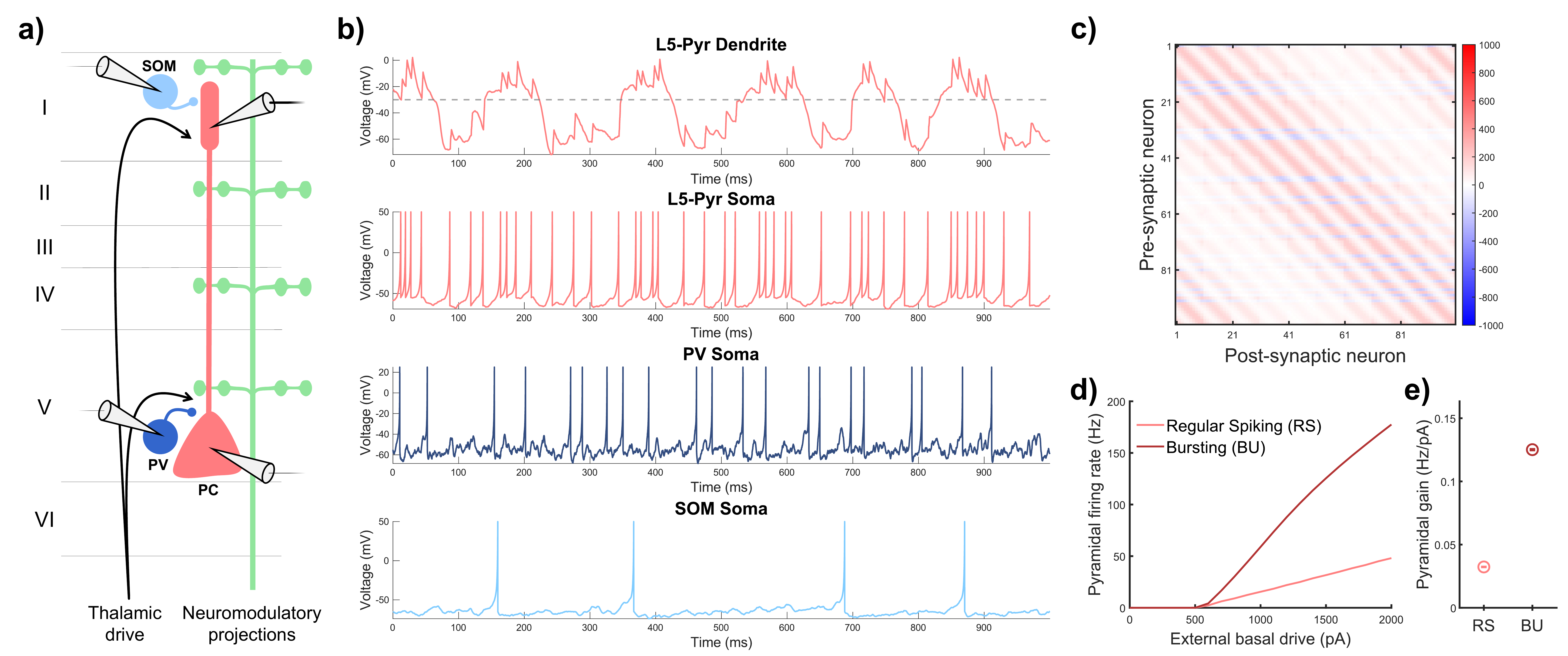}
    \caption{\textbf{Study of layer-5 neurons with PV and SOM inhibition.} \textit{a)} Schematic of the model, illustrating apical activity ($Ca^{2+}$ spikes) and somatic activity (bursts and isolated spikes) in isolation. \textit{b)} Single neuron dynamics of each neuronal type considered in the network: apical and sompatic compartments of layer 5 pyramidal neurons in light red, fast spiking PV interneurons in dark blue and regular spiking SOM interneurons in light blue. Note that if the apical drive on layer 5 pyramidal neurons exeeds the threshold of -30 mV (dashed grey line), the somatic activity exhibit a burst and not isolated spikes. \textit{c)} Initial neuronal connectivity following a gaussian profile. \textit{d)} Pyramidal firing rate comparison for regular spiking (RS) behavior versus a burs-like (BU) behaviour. \textit{e)} Pyramidal gain comparison betuween the regular spiking compared to the bursting regime. Bursts notably increase neuronal gain.}
    \label{fig:diagram}
\end{figure*}

The network comprises a dual-compartment model of pyramidal neurons in layer 5 integrated with fast spiking (FS) PV and regular spiking (RS) dendritic targeting SOM interneurons, providing a simplified representation of the cortical circuitry of layer-5 and the inhibition observed in associative layer 1 (see Figure \ref{fig:diagram}b). The somatic compartments of all cells were modeled using Izhikevich quadratic adaptive integrate-and-fire neurons \cite{izhikevich_simple_2003}. The Izhikevich neuron is described by a two-dimensional system of ODEs:
\begin{equation}
    \begin{aligned}
        \frac{dv_s}{dt} &= \frac{1}{C_s} \left( k_s (v_s - v_r)(v_s - v_t) - u_s + I_s \right), \\
        \frac{du_s}{dt} &= a_s \, \bigl(b_s (v_s - v_r) - u_s\bigr)
    \end{aligned}
\end{equation}
with the reset condition that follows
\begin{equation}
\label{eq:Iz_reset}
\text{if } v_s \geq v_p, \quad
    \begin{cases}
        v_s \leftarrow c_s, \\
        u_s \leftarrow u_s + d_s.
    \end{cases}
\end{equation}
The model equations are formulated in a dimensional form, where the membrane potential dynamics incorporate key biophysical parameters: the resting potential $v_r$, spike threshold $v_t$, spike peak $v_p$, and reset value $c_s$; all measured in mV. Similarly, the input current $I_s = I^{\mathrm{soma},i}_e + \sum_j\sum_k w_{i,j} \; \delta\left( t - t_j^k \right)$, time $t$, and capacitance $C_s$ are expressed in standard biophysical units (pA, ms, and pF, respectively). Four additional dimensionless parameters $k_s$, $a_s$, $b_s$, and $d_s$ shape the dynamical behavior of the model, ie the sharpness of the quadratic non-linearity, the timescale of spike adaptation, the sensitivity of spike adaptation to subthreshold oscillations, and the after spike reset of the adaptation variable; respectively. It is important to note that by effectively tuning these parameters, the model captures biologically plausible spiking behaviors while maintaining computational efficiency, making it a compelling option for neurobiologically realistic simulations \cite{rodriguez-garcia_enhancing_2024, whyte_burst-dependent_2025, gast_neural_2024, munn_neuronal_2023}. Furthermore, these parameters have already been fitted for a wide range of cortical and subcortical neurons based on their spiking types \cite{izhikevich_dynamical_2006}; we summarize the ones used in Table \ref{tab:izhikevich_parameters}.\\
\begin{table*}[t]
\centering
\caption{Izhikevich model parameters for different spiking types}
\label{tab:izhikevich_parameters}
\begin{tabular}{lccccccccc}
\toprule
\textbf{Spiking Type}      & \textbf{$v_p$ (mV)} & \textbf{$C_s$ (pF)} & \textbf{$v_r$ (mV)} & \textbf{$v_t$ (mV)} & \textbf{$k_s$} & \textbf{$a_s$}  & \textbf{$b_s$}  & \textbf{$c_s$ (mV)} & \textbf{$d_s$} \\ \midrule
Regular Spiking (RS)  & 50          & 150        & -75         & -45         & 2.50  & 0.01   & 5.00   & -65       & 250    \\
Bursting (BU)         & 50          & 150        & -75         & -45         & 2.50  & 0.01   & 5.00   & -55       & 150    \\
Fast Spiking (FS)     & 25          & 20         & -55         & -40         & 1.00  & 0.15   & 8.00   & -55       & 200    \\ \bottomrule
\end{tabular}
\end{table*}
The apical compartment of pyramidal neurons is modeled using a generalized integrate-and-fire dynamics that effectively captures the generation of $Ca^{2+}$ plateau potential generated in the apical tufts of layer-5 pyramidal neurons \cite{whyte_burst-dependent_2025,naud_sparse_2018}. Mathematically,
\begin{equation}
\label{eq:apical}
    \begin{aligned}
        \frac{dv_d}{dt} &= \frac{E_{L_d} - v_d}{\tau_d} + \frac{1}{C_d}\Bigl( g_d\, F_{d}(v_d) + w_d + m_d \, P_{\text{BAP}} \, H \left( t - t^{(s)} \right) + I_d \Bigr), \\
        \frac{dw_d}{dt} &= \frac{- w_d + a_d\, (v_d - E_{L_d})}{\tau_{w_d}}
    \end{aligned}   
\end{equation}
where
\begin{equation}
    F_{d}(v_d) = \frac{1}{\left[ 1 + e^{\left(-\frac{v_d+38}{6}\right)} \right]}
\end{equation}
captures the regenerative non-linearity underlying the $Ca^{2+}$ plateau potential, and $H \left( t - t^{(s)} \right)$  represents a square wave function of unitary amplitude that models the BAP, which is delayed by 0.5 ms and lasts for 2 ms. $P_{\text{BAP}}$ denotes the probability that a BAP will occur, capturing the inherent stochasticity observed in experimental studies \cite{short_stochastic_2017}. The parameters $C_d$, $\tau_d$, $g_d$, $m_d$, $\tau_{w_d}$, $a_d$, $I_d = I_e^{\mathrm{dend,pyr}} + \sum_j\sum_k w_{i,j} \; \delta\left( t - t_j^k \right)$ and $E_{L_d}$ correspond to the dendritic capacitance (pF), correspond to the dendritic capacitance (pF), synaptic decay time constant (ms), dendritic coupling conductance (pS), maximum dendritic conductance (pS), adaptation time constant for the dendritic compartment (ms), subthreshold adaptation parameter, dendritic current (pA), and dendritic resting potential (mV); respectively.\\
The compartments are coupled such that somatic sodium spikes trigger backpropagating action potentials (BAPs) probabilistically, with a Bernoulli random variable determined by the soma–apical coupling parameter. Similarly, when the apical membrane potential exceeds –30 mV, $Ca^{2+}$ spikes are generated and conveyed to the soma, switching the firing pattern from isolated spikes to bursts. Taking advantage of the versatility of the Izhikevich model, we can switch a neuron's firing mode simply by adjusting the reset conditions. By increasing the reset voltage $c$ (pushing the neuron closer to threshold) and reducing the spike adaptation $d$ (lowering the post-spike inhibitory effect), the model transitions from regular spiking to bursting. Neuromodulatory signals from the ascending arousal system modulate both the apical drive and the soma-apical coupling, thus fine-tuning the gain of the somatic input-output curve \cite{labarrera_adrenergic_2018,samuels_functional_2008,williams_dendritic_2019,wainstein_gain_2025}.\\
\section{Results and discussion}
Recent empirical findings and cortical models \cite{whyte_burst-dependent_2025,munn_neuronal_2023} guided the design of our model’s dynamic components, which comprise layer 5 bicompartimental pyramidal neurons, fast spiking PV interneurons and regular spiking SOM interneurons. We model each neuronal population with biophysically realistic spiking neurons linked by STDP-governed synapses. This configuration enables the network’s emergent activity to reproduce known neuron-specific firing patterns, bridging computational insights with electrophysiological validation and providing a realistic testbed for investigating how neuromodulation influences spiking dynamics and synaptic plasticity.
\subsection{Apical hysteresis generate calcium plateaus}
\begin{figure*}[ht]
    \centering
    \includegraphics[width=\textwidth]{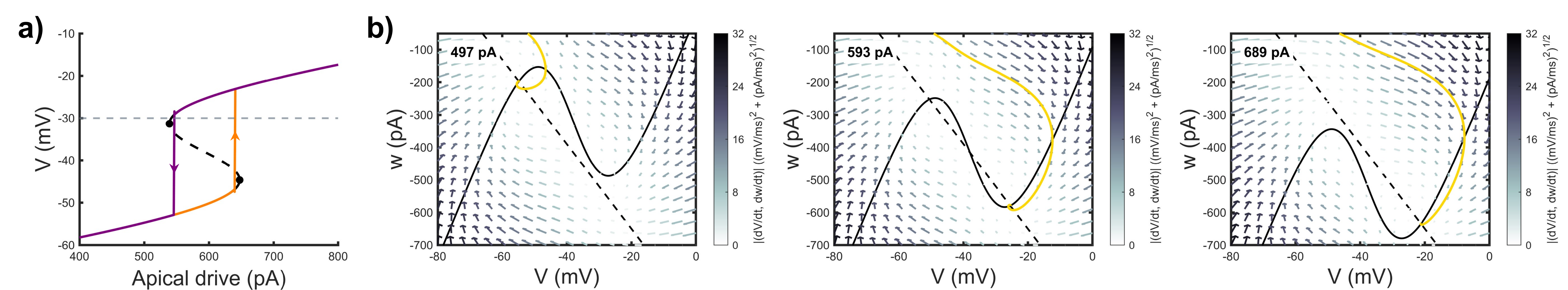}
    \caption{\textbf{Calcium plateaus generation via apical dendritic bistability.} \textit{(a)} Bifurcation diagram of the layer-5 pyramidal neuron apical compartment as the apical drive $I_{d}^{e}$ is varied. A saddle-node bifurcation at $Ie^{(1)}_{d}=538.91$pA gives rise to a stable plateau potential coexisting with the resting state. At a second saddle-node bifurcation  $Ie^{(2)}_{d}=647.37$pA the resting state disappears and the plateau potential becomes the unique globally attracting solution. \textit{(b)} Phase-plane views of the apical compartment for three regimes of the external current: $I_{\rm ext}<Ie^{(1)}_{d}$ (left),  $Ie^{(1)}_{d}<I_{\rm ext}<Ie^{(2)}_{d}$ (center), and $I_{\rm ext}>Ie^{(2)}_{d}$ (right). Black curves are the voltage and recovery-variable nullclines, arrows show the vector field, and colored trajectories illustrate the system’s response.}
    \label{fig:hysteresis}
\end{figure*}

To effectively model BAC firing with computational efficiency, we utilize a novel dual-compartmental approach proposed by Whyte et al. (2025) \cite{whyte_burst-dependent_2025}. The model consists of a somatic compartment implemented via an Izhikevich adaptive quadratic integrate-and-fire neuron \cite{izhikevich_dynamical_2006,izhikevich_simple_2003}, and an apical compartment represented by a generalized integrate-and-fire model of $Ca^{2+}$ plateau dynamics \cite{whyte_burst-dependent_2025,naud_sparse_2018}. Importantly, the two compartments are linked such that somatic sodium spikes initiate BAPs probabilistically through a Bernoulli process controlled by a soma-apical coupling parameter, aligning with neurophysiological observations \cite{short_stochastic_2017}. Consequently, when sufficient apical drive coincides with this mechanism, it elicits a $Ca^{2+}$ plateau potential exceeding –30 mV, transitioning the neuron from regular spiking to bursting by modifying the somatic reset conditions. This bursting regime exhibits a higher input-output gain than isolated spikes (see Figure \ref{fig:diagram}d and e), thereby enhancing the responsiveness of the pyramidal neuron.\\
The generation of $Ca^{2+}$ plateaus arises from a hysteresis bifurcation occurring within the apical compartment. To understand the factors influencing the system's transition between stable points, we analyze this compartment using dynamical systems theory. By calculating the nullclines ($\dot{v_d} = \dot{w_d} = 0$), we obtain:
\begin{equation}
\begin{aligned}
a_d \left(v - E_{L_d}\right) &= \frac{C_d}{\tau_d} \left(E_{L_d} - v\right) - g_d F_{d}(v) - Ie_d \
w &= -\frac{C_d}{\tau_d} \left(E_{L_d} - v\right) - g_d F_{d}(v) - Ie_d
\end{aligned}\end{equation}
This analysis reveals a region of bistability across a specific range of external dendritic currents ($Ie_d$). Hence, the neuron can settle into one of two distinct stable states depending on the direction and magnitude of changes in $Ie_d$ (see Figure \ref{fig:hysteresis}a, $Ie^{(1)}_d = 647.37$~pA and $Ie^{(2)}_d = 538.91$~pA). By explicitly rescaling the dendritic current as 
\begin{equation}
Ie_d = m_d \, P_{\text{BAP}} \, H \left( t - t^{(s)} \right) + I_d,
\end{equation}
we illustrate that BAP-triggered currents and dendritic inputs together modulate the neuron's stability. Essentially, this factors switch the system between stable state, one of which is above the -30 mV threshold propmpting the generation of $Ca^{2+}$ plateaus and hence switching pyramidal cell spiking to bursts. Furthermore, knowing that burst-like behavior increase the neuronal gain (Figure \ref{fig:diagram}d), one can think of a system that by fine-tuning this parameters dynamically is able to adapt neuronal gain on demand. In fact, this is what neuromodulators are thought to do within biological circuits \cite{rodriguez-garcia_enhancing_2024, mei_improving_2025}, and specifically those associated with the ascending arousal system have been particularly linked to this gain-modulated mechanisms \cite{shine_neuromodulatory_2023, munn_thalamocortical_2023, whyte_burst-dependent_2025, wainstein_gain_2025}.
\subsection{Somatic-to-apical coupling and dendritic drive modulate the pyramidal gain}
\begin{figure*}[ht]
    \centering
    \includegraphics[width=\textwidth]{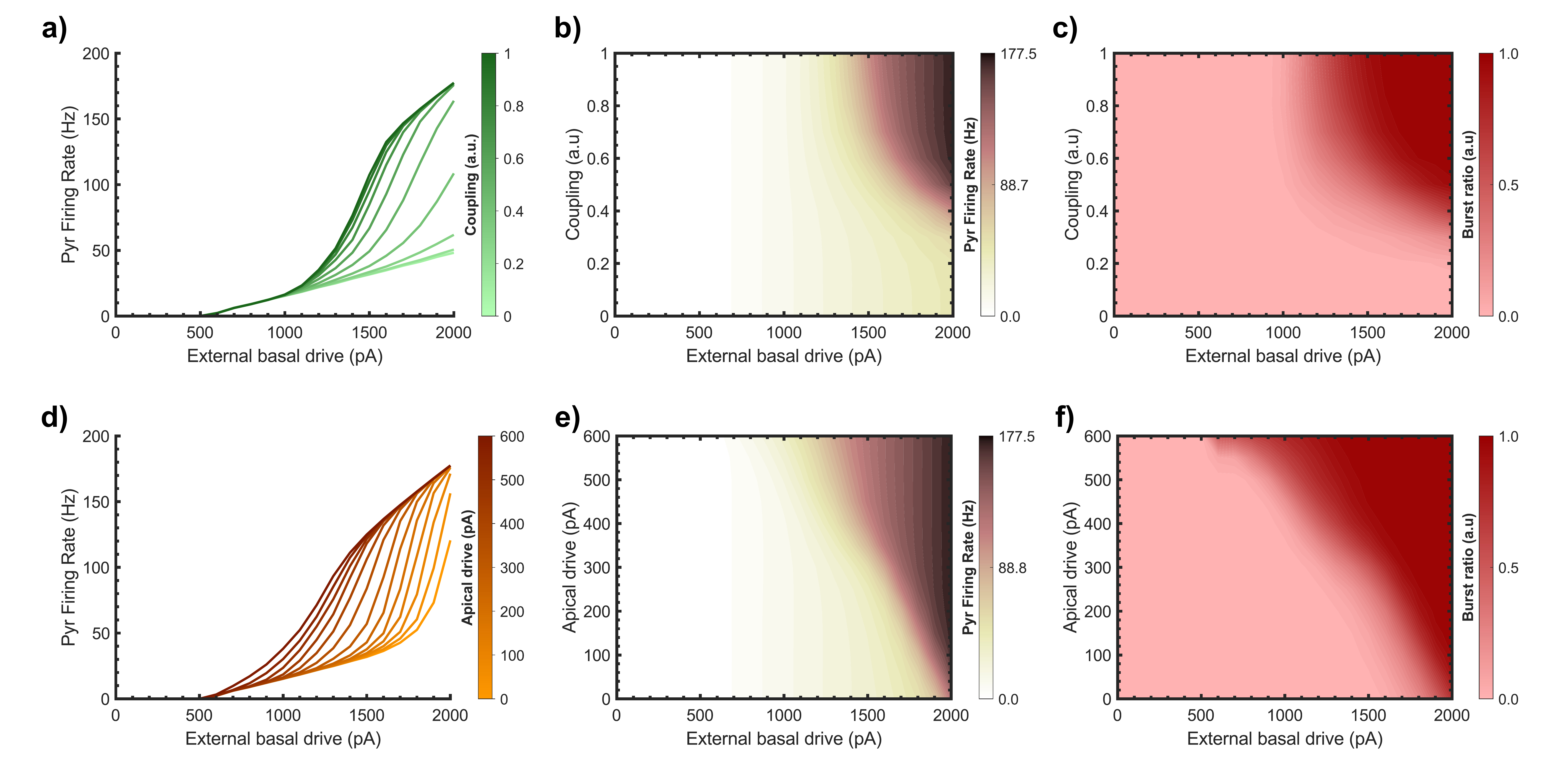}
    \caption{\textbf{Modulation of somatic responses by apical-somatic coupling and apical dendritic drive in layer 5 pyramidal neurons.} \textit{Effect of apical-somatic coupling:} Simulations were performed with a OU apical drive of 400~pA, sufficient to shift the stability point of the dendritic nonlinearity via backpropagating action potentials (BAPs). \textit{a)} Somatic firing rate curves for varying apical-somatic coupling strengths show that increased coupling amplifies the response to basal input. \textit{b)} Heat map of the somatic firing rate as a function of basal drive and coupling strength. \textit{c)} Burst ratio in the same parameter space, revealing that bursting emerges predominantly at high levels of both basal drive and coupling. Light red denotes a regular spiking regime, while dark red indicates a fully bursting regime. \textit{Effect of apical dendritic drive:} In these simulations, neurons were fully coupled (maximal apical-somatic coupling). \textit{d)} Somatic firing rate curves across apical drive levels demonstrate gain modulation analogous to that observed with increasing coupling. \textit{e)} Heat map of somatic firing rate as a function of basal and apical drive. \textit{f)} Burst ratio across the same input space, indicating that bursting arises when both apical and basal inputs are strong. All simulations were conducted for 1000 ms and 100 ms of initial delay over 1000 trials, each with independent random-noise realizations.}
    \label{fig:neuron_coup_dend}
\end{figure*}

To examine the effects of neuromodulatory processes in our model, we altered two critical parameters in which neuromodulatory processes are involved affecting pyramidal firing: the compartmental coupling and the apical dendritic drive. Panel a of Figure \ref{fig:neuron_coup_dend} shows that strengthening somatic-apical coupling boosts the input-output response of an isolated layer 5 pyramidal neuron and panel b presents the same effect as a continuous colormap. Notably, as coupling increases, the neuron’s firing rate rises more steeply for the same input, reflecting a higher gain. Furthermore, for high-coupling, the gain increases in a supra-linear manner with respect to the linear manner that it does for low-coupling. This effect can be understood through the backpropagating action potentials, i.e., stronger coupling increases the probability that each somatic spike will invade the apical compartment, increasing the likelihood to trigger $Ca^{2+}$ spikes and elevating the overall firing rate through a positive feedback loop by bursting activity (Figure \ref{fig:neuron_coup_dend}c). This steep increase in firing rate is indicative of heightened neuronal gain, suggesting that neuromodulatory signals adjusting soma-apical coupling effectively regulate gain in layer 5 pyramidal neurons, aligning with recent findings \cite{labarrera_adrenergic_2018, whyte_burst-dependent_2025}. Specific neuromodulatory processes related with arousal mediate this parameter; for instance, noradrenaline released from the locus coeruleus acting on $\alpha$2A receptors induces closure of HCN channels along the apical shaft \cite{samuels_functional_2008}, and diffusely projecting thalamic inputs target oblique dendrites \cite{suzuki_general_2020} -- both of which have also been discussed in other modeling studies \cite{munn_neuronal_2023,munn_thalamocortical_2023}.\\
Furthermore, neuromodulatory processes in the brain can influence both the correlation structure and the magnitude of apical input received by pyramidal neurons. Non-specific thalamic projections \cite{jones_viewpoint_1998, sherman_thalamocortical_2012}, as well as input from the dorsal raphe nucleus and from the locus coeruleus, can modulate the timing and strength of signals reaching the apical dendrites \cite{samuels_functional_2008, zhang_gain_2005}. In addition, ACh enhances apical excitability via depolarization mediated by M1 muscarinic receptors, thereby increasing the neuron's sensitivity to top-down signals \cite{williams_dendritic_2019}. An increase in apical input, regardless of its neuromodulatory origin, can enhance dendritic excitability and boost neuronal gain. In this study, we focus on the effects of increased apical input as a mechanism for gain modulation. Previous computational work has also explored how the correlation of apical inputs contributes to maximizing information processing in pyramidal neurons \cite{munn_thalamocortical_2023, munn_neuronal_2023}. Panels d and e in Figure \ref{fig:neuron_coup_dend} show how increasing apical drive increases neuronal gain. Unlike the sharp nonlinearity induced by increased coupling, the gain modulation here is smoother and more graded, reflecting the direct depolarizing influence of distal dendritic input. Notably, panel \ref{fig:neuron_coup_dend}f reveals that bursting becomes prominent when both apical and basal inputs are strong, reinforcing the view that bursting acts as a coincidence detector for converging bottom-up and top-down signals, which has computational implications in how predictive coding processes and backpropagation learning may be happening in the brain \cite{sacramento_dendritic_2018, greedy_single-phase_2022, aizenbud_neural_2025}.
\subsection{Dendritic inhibition and somatic inhibition fine-tune and gate pyramidal activity}
\begin{figure*}[ht]
    \centering
    \includegraphics[width=\textwidth]{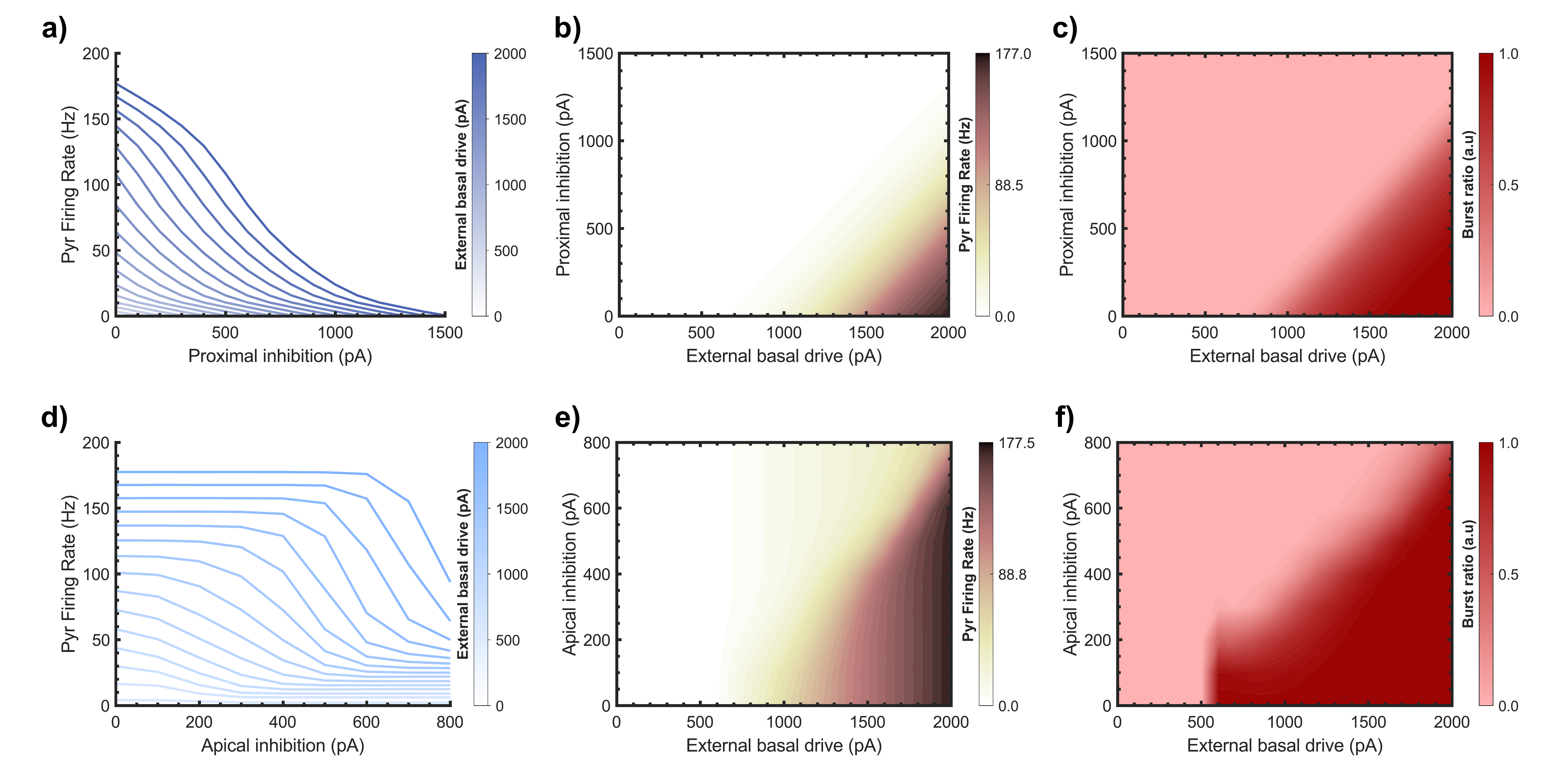}
    \caption{\textbf{Suppression of somatic responses by inhibitory control at proximal and apical dendritic compartments in layer-5 pyramidal neurons.} \textit{Effect of proximal (somatic) inhibition:} Simulations were performed with a OU apical drive of 500~pA and fully coupled neurons, to ensure simulating BAC firing in the system. \textit{a)} Somatic firing rate curves across increasing levels of proximal inhibition demonstrate a progressive suppression of firing, with stronger inhibition requiring higher basal drive to elicit output. \textit{b)} Heat map showing the firing rate as a function of basal drive and proximal inhibition strength. \textit{c)} Burst ratio in the same parameter space reveals that bursting is abolished under moderate-to-strong proximal inhibition, confirming its dominant role in regulating overall output. \textit{Effect of dendritic inhibition:}Simulations were performed with fully connected neurons and a strong apical drive (800~pA) to show the inhibitory effects clearly. \textit{d)} Somatic firing rate curves under increasing dendritic (apical) inhibition show that, unlike proximal inhibition, firing remains relatively preserved, though bursting is strongly suppressed. \textit{e)} Heat map of firing rate as a function of basal drive and dendritic inhibition illustrates a milder effect on output firing. \textit{f)} Burst ratio across the same space reveals that dendritic inhibition primarily reduces bursting without significantly affecting regular spiking, demonstrating its selective influence on dendritic excitability and $Ca^{2+}$ spike initiation. All simulations were conducted for 1000 ms and 100 ms of initial delay over 1000 trials, each with independent random-noise realizations.}
    \label{fig:neuron_inh}
\end{figure*}

Dendritic and somatic inhibition independently shape the firing response of an isolated layer 5 pyramidal neuron. In Figure \ref{fig:neuron_inh}a–c, we examine the effect of proximal (somatic) inhibition by applying varying levels of inhibitory current to the soma while keeping mean apical drive constant at 500~pA. Increasing proximal inhibition systematically suppresses the firing rate, indicating a higher threshold for spike initiation and a global reduction in neuronal excitability, i.e. shunting inhibition, replicating neurophysiology  experimental findings \cite{strack_biological_2013}. Although the Izhikevich model used here does not impose a hard upper bound on firing rates \cite{strack_biological_2013}. Notably, neuromodulators such as acetylcholine (ACh) can modulate PV interneuron excitability, adjusting inhibitory efficacy and thereby shunting pyramidal neurons in a behaviorally state-dependent manner \cite{gast_neural_2024, colangelo_cellular_2019}. This may enable segregation during learning by turning off particular information processing pathways \cite{shine_neuromodulatory_2023}.\\
In Figure \ref{fig:neuron_inh}d–f, we explore the role of dendritic inhibition by applying increasing levels of inhibitory current to the apical compartment, while holding a strong mean dendritic excitatory drive constant at 800~pA. As dendritic inhibition increases, the ability of the apical dendrite to generate $Ca^{2+}$ spikes is progressively suppressed, resulting in a significant decrease in burst firing. Unlike somatic inhibition, dendritic inhibition modulates gain in a more graded and compartment-specific fashion, transitioning the neuron from a high-gain, bursting regime to a low-gain, regular spiking mode. This effect reflects the role of SOM interneurons, which preferentially target distal dendrites and exert a strong influence on nonlinear dendritic integration \cite{pouille_contribution_2013}. These interneurons effectively gate the influence of top-down and contextual inputs arriving at the apical tuft, enabling selective control over burst-dependent signaling.
\subsection{Neuromodulation of the pyramidal gain modulates network plasticity update rates in STDP learning}
\begin{figure}[htbp!]
    \centering
    \includegraphics[width=\columnwidth]{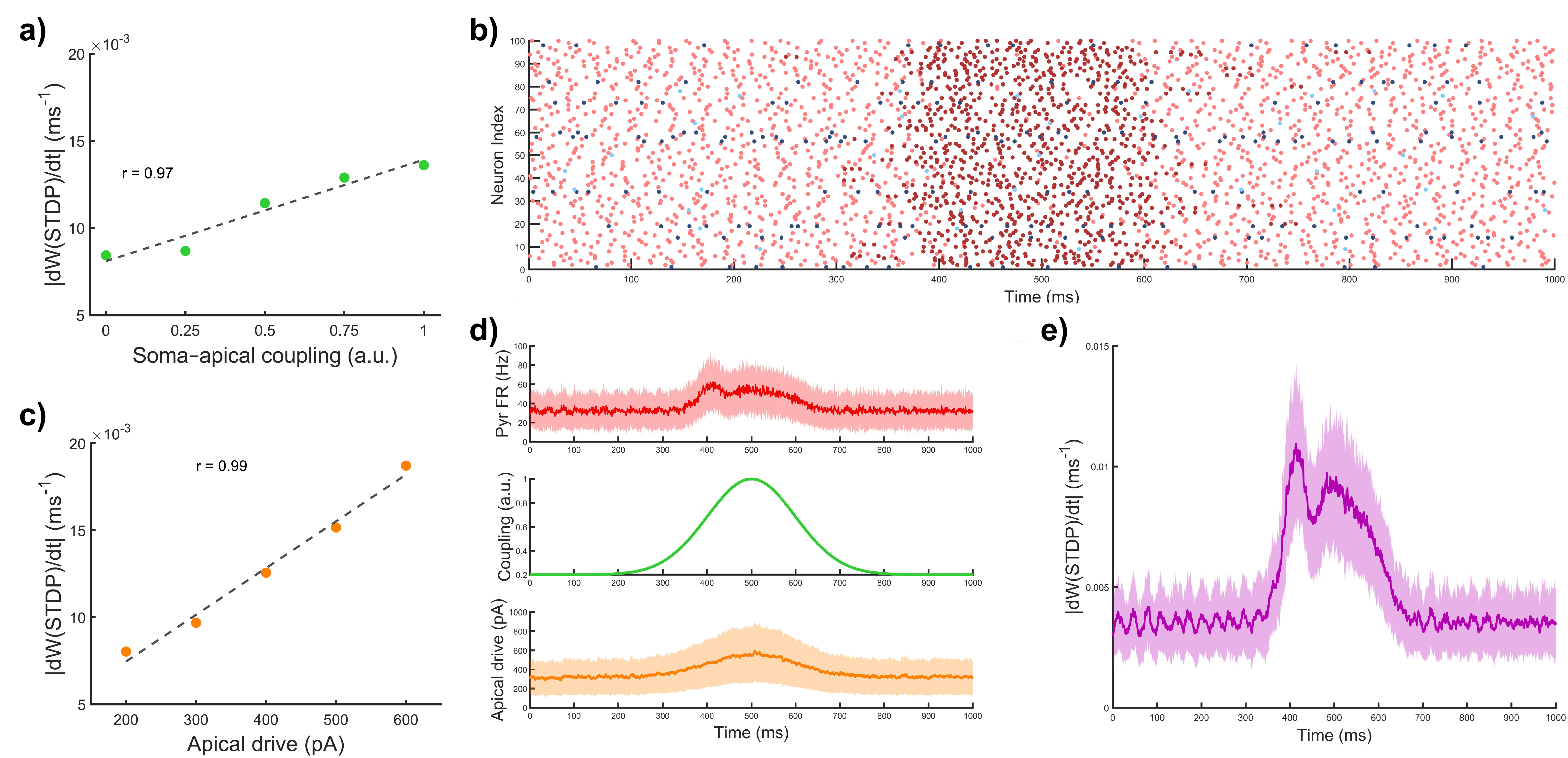}
    \caption{\textbf{Gain neuromodulation enhances STDP update rates:} Simulations were made with a 100~ms delay, 2000~pA of basal drive, and a drive to SOM inhibitory neurons of 600~pA and none to PV interneurons. \textit{a)} In this case, we used 450~pA of apical drive and sweeped the coupling paramenter, showing  the increase in neuronal gain by the soma-apical coupling increases the plasticity rate of the pyramidal neurons. \textit{c)} In this case we considered a fully-coupled pyramidal neuron and sweeped the apical current, illustrating how an increased dendritic drive increase more notably the dendritic drive. Error bars are not presented as the dot size is bigger than their magnitude. \textbf{Dynamical gain modulation:} Simulations were carried with 500~ms delay, 1500~pA of basal drive, and a drive to SOM inhibitory neurons of 600~pA and none to PV interneurons. \textit{b)} Network rasterplot showing the different neuronal types: pyramidal isolated spikes in light red, pyramidal bursts in dark red, SOM interneurons in light blue and PV interneurons in dark blue. \textit{d)} Top: pyramidal firing rate evolution over time. Middle: Coupling evolution over time, Bottom: apical drive evolution over time. \textit{e)} Evolution of the rate of synaptic changes over time, showing that phasic neuromodulation increase the rate of synaptic changes providing flexibility on the network's connectivity. All simulations were conducted for 1000 ms over 100 trials, each with independent random-noise realizations.}
    \label{fig:gain_neuromodulation}
\end{figure}

Motivated by the implications of neuromodulation during learning, we examine how gain variations alter synaptic plasticity. We hypothesize that if synaptic plasticity depends on the spiking behavior of neurons, then synaptic updates occur more rapidly at higher firing rates. To test this, we systematically vary the soma-apical coupling and dendritic apical drive, measuring their effects on the rate of synaptic weight changes under STDP. Panels a and c of Figure~\ref{fig:gain_neuromodulation} show that increasing either soma–apical coupling (panel a) or apical drive (panel c) results in a near‐linear rise in the absolute STDP update rate. In panel a, with the mean basal drive held constant at 450~pA and the coupling parameter swept from 0 to 1, we observed a strong positive correlation ($r=0.97$) between coupling strength and plasticity rate. In panel (c), under full coupling, varying the apical current from 200 to 600~pA similarly yielded a robust linear increase ($r=0.97$). These findings demonstrate that both coupling‐ and drive‐mediated mechanisms can effectively tune pyramidal gain and accelerate learning. Notably, the slope for apical drive exceeds that for pyramidal coupling, indicating a stronger influence of apical drive on synaptic update dynamics.\\
From a functional perspective, neuromodulators such as ACh and NA can independently or jointly adjust the apical drive and soma–apical coupling, effectively toggling the network between high-gain, flexible states (in which spike rates and STDP updates are elevated) and low-gain, stable states (with reduced spiking and slower weight changes), aligning with theoretical studies \cite{wainstein_gain_2025, munn_neuronal_2023, munn_thalamocortical_2023, whyte_burst-dependent_2025}. This capacity for dynamic gain modulation is vital for adaptive behavior, as it allows the system to balance exploratory, rapidly learning phases with more conservative, stable phases. Future work will analyze the network’s energy states in greater detail to substantiate this point, and will employ simple tasks to test how NA- and ACh-like signals can invoke flexible or stable processing modes on demand.\\
To illustrate a behaviorally plausible scenario, we examine a brief neuromodulator driven increase in pyramidal gain that can arise during periods of perceptual uncertainty \cite{samuels_functional_2008, wainstein_gain_2025}. We hypothesize that a phasic burst of NA from the locus coeruleus simultaneously augments recurrent coupling and apical dendritic drive, accelerating spike timing-dependent plasticity and supporting rapid integration and flexible encoding. To test this idea we applied a Gaussian modulation to both the mean apical drive and the pyramidal coupling (Figure \ref{fig:gain_neuromodulation}d), mimicking integration of information under uncertainty \cite{wainstein_gain_2025}. The top panel of Figure \ref{fig:gain_neuromodulation}d shows that this modulation transiently raises pyramidal neuron gain, and the resulting increase in plasticity update rates enables real time adaptation, shown in Figure \ref{fig:neuron_coup_dend}. Interestingly, this phasic surge in plasticity decouples every synaptic weight into a fast-slow regime (see Supplementary \ref{app:fast-slow}), where a fast component that quickly absorbs new sensory evidence and a slow component that preserves existing knowledge, creating a fast slow regime that promotes integration with minimal interference \cite{hinton1987}.\\
\section{Outlook}
Our model reproduces key neurophysiological observations, including how apical drive boosts dendritic gain and how compartment specific somatostatin and parvalbumin inhibition respectively dampen and raise pyramidal excitability, grounding our simulations in experimental data \cite{pouille_contribution_2013,larkum_top-down_2004}. Building on this foundation, we show that a transient increase in pyramidal gain rescales the step size of spike timing dependent plasticity, complementing earlier work demonstrating that gain modulation reshapes the cortical energy landscape \cite{munn_neuronal_2023, munn_thalamocortical_2023, wainstein_gain_2025, whyte_burst-dependent_2025}. By altering the learning rate directly at the synapse, neuromodulatory signals can thus enact rapid shifts in network plasticity without invoking separate third‐factor rules.\\
Crucially, we further show that phasic plasticity changes causes each synaptic weight to decompose into fast and slow components -- one that rapidly incorporates new sensory evidence and another that preserves established connections (see Supplementary \ref{app:fast-slow}). Such a fast‐slow weight decomposition offers a pathway for enhancing continual learning and suggests novel designs for neuromorphic attention pipelines that toggle between rapid adaptation and stable processing.
%%% REFERENCES %%%
\newpage
\bibliography{sample}

%%% APPENDIX %%%
\newpage
\appendix
\section*{APPENDIX}
\section{Plasticity decomposition in fast–slow weights}
\label{app:fast-slow}
In this Appendix we derive the decomposition of the synaptic update into “slow” (tonic) and “fast” (phasic) components. We begin by writing the continuous‐time plasticity rule as
\begin{equation}
\frac{dw}{dt} = \Gamma\bigl(w(t),t\bigr),
\end{equation}
where $\Gamma$ denotes the instantaneous rate of change of the weight $w(t)$. Discretizing with learning rate $\eta$ and time step $\Delta t$ gives
\begin{equation}
w(t+\Delta t)
= w(t) + \eta\,\Gamma\bigl(w(t),t\bigr)\,\Delta t.
\end{equation}
To separate tonic and phasic components, we introduce a low‐pass filtered version of $\Gamma$, denoted $\Gamma_0(t)$, with time constant $\tau$:
\begin{equation}
\tau\,\frac{d\Gamma_0}{dt} = -\,\Gamma_0(t) + \Gamma\bigl(w(t),t\bigr),
\end{equation}
so that
\begin{equation}
\Gamma_0(t)
= \frac{1}{\tau}
  \int_{-\infty}^t 
    e^{-(t-s)/\tau}\,
    \Gamma\bigl(w(s),s\bigr)\,ds,
\end{equation}
captures the slow (tonic) component. We then decompose the update rate as
\begin{equation}
\Gamma\bigl(w(t),t\bigr)
= \Gamma_0(t) + \bigl[\Gamma\bigl(w(t),t\bigr)-\Gamma_0(t)\bigr],
\end{equation}
where the second term is the fast (phasic) fluctuation. Substituting into the discrete update,
\begin{equation}
w(t+\Delta t)
= w(t)
  + \eta\,
    \Bigl[
      \Gamma_0(t)
      + \bigl(\Gamma(w,t)-\Gamma_0(t)\bigr)
    \Bigr]
    \Delta t.
\end{equation}
Finally, defining
\begin{equation}
w(t)=w_{\rm slow}(t)+w_{\rm fast}(t),
\end{equation}
and imposing
\begin{align}
\frac{d\,w_{\rm slow}}{dt} &= \eta\,\Gamma_0(t),\\
\frac{d\,w_{\rm fast}}{dt} &= \eta\,\bigl[\Gamma(w,t)-\Gamma_0(t)\bigr],
\end{align}
we obtain two coupled dynamics whose sum reproduces the full update \footnote{At first-order approximation (Euler integration).}. The variable $w_{\rm slow}$ integrates the slowly varying component of the plasticity signal, encoding stable, long‐term changes, while $w_{\rm fast}$ captures rapid adjustments to fluctuations in $\Gamma$, allowing quick adaptation without overwriting the tonic memory \cite{hinton1987}.
\section{Supporting information}
Here we provide additional plots supporting the results already discussed in the main text, highlighting how variations in apical somatic coupling affect the network of pyramidal neurons, and apical dendritic drive and inhibitory inputs modulate layer 5 pyramidal neuron firing dynamics (input-output curves).
\begin{figure*}[ht]
    \centering
    \includegraphics[width=\textwidth]{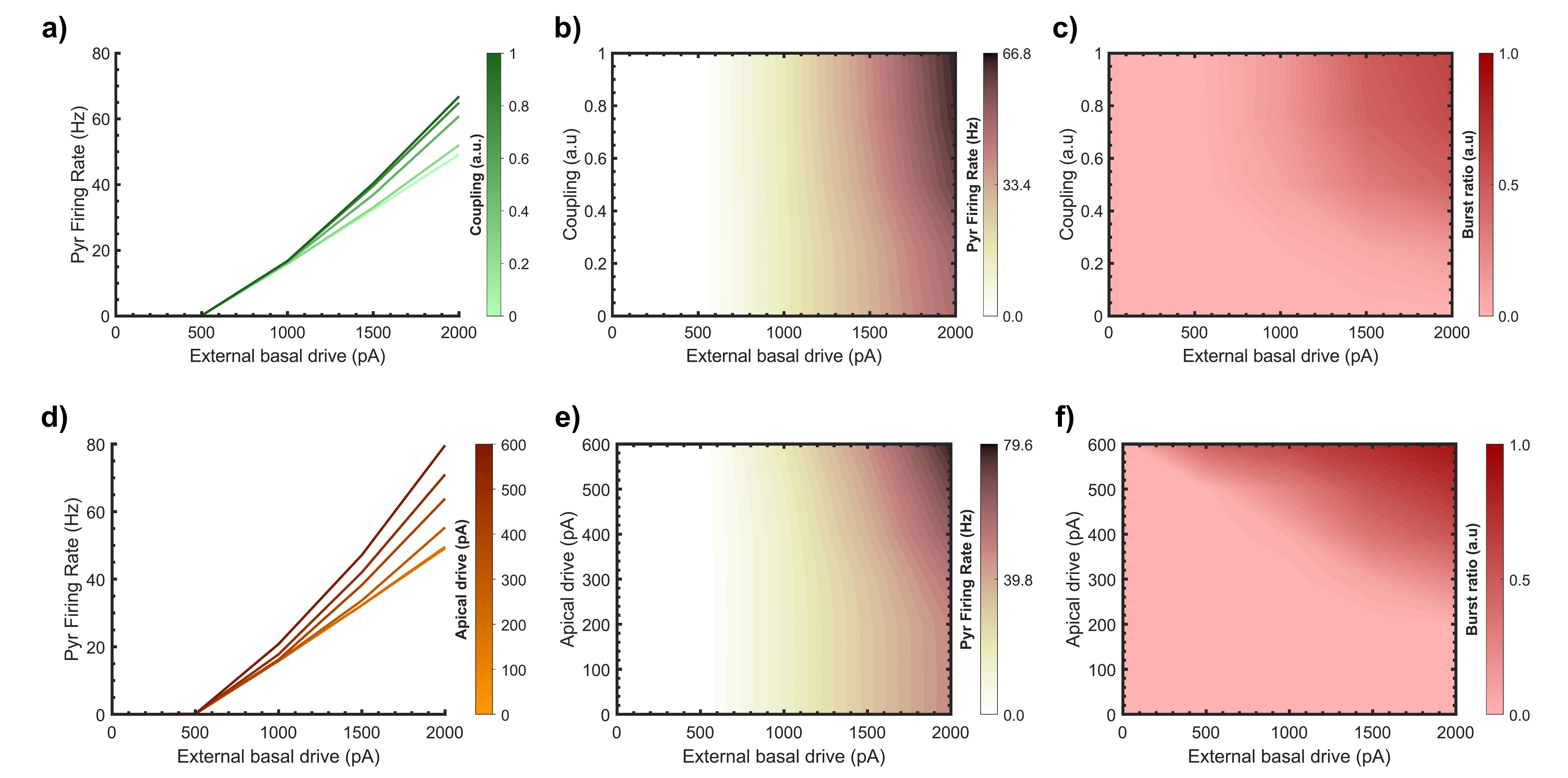}
    \caption{\textbf{Modulation of somatic responses by apical-somatic coupling and apical dendritic drive in a network of layer 5 pyramidal neurons.} \textit{Effect of apical-somatic coupling:} Simulations were performed with a OU apical drive of 450~pA, 600~pA to SOM neurons and 0~pA to PV interneurons. \textit{a)} Somatic firing rate curves for varying apical-somatic coupling strengths show that increased coupling amplifies the response to basal input. \textit{b)} Heat map of the somatic firing rate as a function of basal drive and coupling strength. \textit{c)} Burst ratio in the same parameter space, revealing that bursting emerges predominantly at high levels of both basal drive and coupling. Light red denotes a regular spiking regime, while dark red indicates a fully bursting regime. \textit{Effect of apical dendritic drive:} In these simulations, neurons were fully coupled (maximal apical-somatic coupling). \textit{d)} Somatic firing rate curves across apical drive levels demonstrate gain modulation analogous to that observed with increasing coupling. \textit{e)} Heat map of somatic firing rate as a function of basal and apical drive. \textit{f)} Burst ratio across the same input space, indicating that bursting arises when both apical and basal inputs are strong. All simulations were conducted for 1000 ms and 100 ms of initial delay over 1000 trials, each with independent random-noise realizations.}
    \label{fig:app_net_coup_dend}
\end{figure*}

\begin{figure*}[ht]
    \centering
    \includegraphics[width=\textwidth]{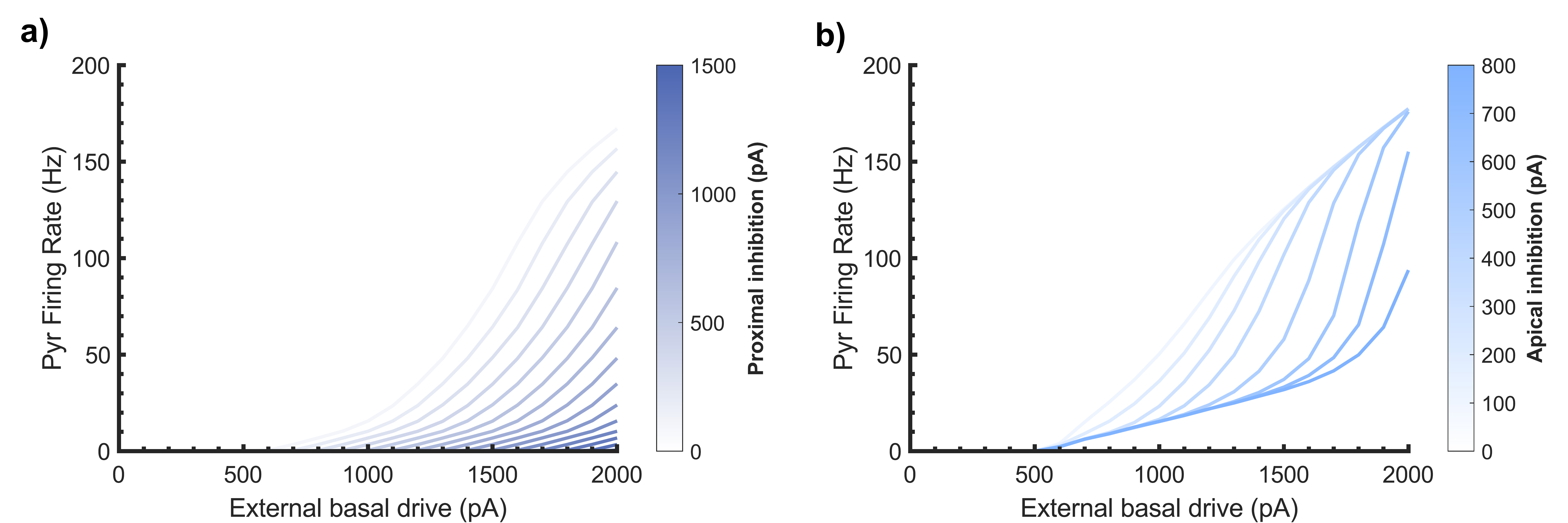}
    \caption{\textbf{Inhibition of layer 5 pyramidal neurons.} \textit{a)} Effect of proximal (somatic) inhibition on the input-output curve of pyramidal neurons ('shunting effect'). \textit{b)} Effect of apical (dendritic) inhibition on the input-output curve of pyramidal neurons ('gain modulation'). Simulations were conducted identically as in Figure \ref{fig:neuron_inh}.}
    \label{fig:app_inh}
\end{figure*}

\end{document}